\documentclass[useAMS,usenatbib]{mn2e}
\usepackage{amstext}
\usepackage{amssymb}
\usepackage{aas_macros}
\usepackage{amsmath}
\usepackage{graphicx}
\usepackage{txfonts}
\usepackage{mathtools,revsymb}
\usepackage{booktabs,multirow,calc,xspace,rotating}
\usepackage{hyperref}

\newcommand{\partialderiv}[2]{\frac{\partial #1}{\partial #2}}
\newcommand{\diverg}[1]{\nabla \cdot #1}

\def \Msun{M$_\odot$}

\def \Pimin{\Pi_{\rm min}}
\def \Pimax{\Pi_{\rm max}}
\def \epsnet{\epsilon_{\rm net}}
\def \epsnuc{\epsilon_{\rm nuc}}
\def \intM{\int_{0}^{M_{ad}}}
\def \Wnuc{W_{\rm nuc}}
\def \Wrad{W_{\rm rad}}
\def \Wgrav{W_{\rm grav}}

\def \ttherm{t_{\rm therm}}
\def \tmode{\tau_{\rm mode}}
\def \tconv{\tau_{L}}

\title[Stability of Massive Main Sequence Stars]
	{The Stability of Massive Main Sequence Stars as a Function of Metallicity}
	
\author[Shiode, et~al.]{Joshua~H.~Shiode,$^{1}$\thanks{E-mail:
    jshiode@astro.berkeley.edu} Eliot~Quataert,$^{1}$ 
   	Phil~Arras$^{2}$\\
$^{1}$Department of Astronomy, University of California, Berkeley, CA 94720-3411, USA\\
$^{2}$Department of Astronomy, University of Virginia, P.O. Box 400325, Charlottesville, VA 22904-4325, USA\\
}

\begin{document}
\date{Accepted  . Received   ; in original form  }
\pagerange{\pageref{firstpage}--\pageref{lastpage}} \pubyear{2011}
\maketitle
\label{firstpage}

\begin{abstract}

We investigate the pulsational stability of massive ($M \gtrsim 120$ \Msun{}) main sequence stars of a range of metallicities, including primordial, Population III stars.  We include a formulation of convective damping motivated by numerical simulations of the interaction between convection and periodic shear flows.  We find that convective viscosity is likely strong enough to stabilize radial pulsations whenever nuclear-burning (the $\epsilon$-mechanism) is the dominant source of driving. This suggests that massive main sequence stars with $Z \lesssim 2 \times 10^{-3}$ are pulsationally stable and are unlikely to experience pulsation-driven mass loss on the main sequence. These conclusions are, however, sensitive to the form of the convective viscosity and highlight the need for further high-resolution simulations of the convection-oscillation interaction. For more metal-rich stars ($Z \gtrsim 2 \times 10^{-3}$), the dominant pulsational driving arises due to the $\kappa$-mechanism arising from the iron-bump in opacity and is strong enough to overcome convective damping. Our results highlight that even for oscillations with periods a few orders of magnitude shorter than the outer convective turnover time, the ``frozen-in'' approximation for the convection-oscillation interaction is inappropriate, and convective damping should be taken into account when assessing mode stability.


\end{abstract}



\section{Introduction} \label{sec:intro}

Several rounds of investigations over the course of the last seventy years have shown that massive main sequence stars above a critical mass near 100 \Msun{} are vibrationally unstable in their fundamental radial mode \citep[e.g.][]{l41,sh59,ss70,z70}. In these stars, the low density contrast between the stellar core and envelope allows the fundamental mode to reach large amplitude near the core, where it may readily couple to the highly temperature sensitive nuclear energy generation (the $\epsilon$-mechanism). This results in linear instability with growth times shorter than the stellar evolutionary time. The resulting non-linear evolution of this instability, which several authors proposed could lead to significant pulsation-driven mass loss, remains an unsolved problem despite considerable effort \citep[e.g.][]{a70, z70, p73}.




The introduction of OPAL opacities in stellar models in the early nineties \citep{ri92} enhanced the linear instability of massive stars of approximately solar metallicity. \citeauthor{gkc+99} \citeyearpar{gk93,gkc+99} found that the enhanced opacity in the stellar envelope due to transitions in heavy element isotopes can produce strong radiative ($\kappa$-mechanism) and strange-mode driving. Today, these mechanisms are generally believed to be the most important sources of fundamental mode instability in massive solar-metallicity stars \citep{g05}.

In the investigations to date, convection has been treated in the ``frozen-in'' approximation, in which the interaction between pulsation and convection is ignored entirely. The validity of this approximation is unclear and needs to be explicitly quantified. The treatment of convection-pulsation interaction as an enhanced viscosity has a long history rooted in studies of tidal Q factors and the solar 5-minute oscillations \citep[e.g.][]{gn77,z89}. Only recently, however, have these theoretical estimates been calibrated with numerical experiments. In particular, \citeauthor{psr+09} \citeyearpar{psr+09,ps11} have shown that for oscillation timescales of the same order as the outer convective turnover time, the interaction is well represented by an anisotropic viscosity which scales linearly with the ratio of the oscillation and outer convective turnover times \citep[as argued for by][]{z89}. On timescales shorter than about one-third of the outer turnover time, these authors argue for a quadratic scaling in accordance with \citet{gn77}.

In this paper we reconsider the linear stability of radial oscillations in massive stars, including primordial Population III stars.  For primordial stars, the lack of metals excludes opacity-driven instabilities in the envelope (i.e., $\kappa$-mechanism and strange modes). The driving found previously for Population III stars above about 120 \Msun{} relies on nuclear-driving in the convective stellar core ($\epsilon$-mechanism), with resulting growth times that are much longer than for opacity-driven modes \citep{bhw01,su11}. It is thus particularly important to check the effect of convective damping on these weakly-driven modes. We also extend our investigation to higher metallicities for $\sim 100$ \Msun{} models to examine the effect of convective damping on the stability of massive stars of a range of metallicities. 

We begin by describing the equilibrium stellar models used in this work in Section \ref{sec:models}. In Section \ref{sec:pulse}, we describe our quasi-adiabatic linear stability analysis, highlighting the important contributions to mode driving and damping. We then describe our primary results in Section \ref{sec:res} and discuss their implications in Section \ref{sec:disc}.


\section{Equilibrium Models} \label{sec:models}

We have computed evolutionary sequences for equilibrium stellar models with initially primordial composition (i.e., Population III stars) using the MESA \texttt{star} stellar evolution code \citep{mesa11}\footnote{http://mesa.sourceforge.net/}. We have also computed model sequences for an initial mass of 120 \Msun{}, and a range of metallicities from $Z = 2 \times 10^{-6}$ to solar (which we take to be $Z = 0.02$). All models are non-rotating and have mass loss turned off. The former approximation is made for simplicity, while the latter is justified for Population III stars because the line-driving mechanism for massive star winds relies on the presence of metal-line opacity in the stellar atmosphere \citep[e.g.,][]{lc99,cak75}. For the higher metallicity models, since we are interested only in the stability of massive stars on the main sequence, the effects of mass loss are not critical.


We use the standard mixing length prescription of \citet{MLT} and a mixing length of $\sim 1.6$ pressure scale heights for the convective mixing. To examine the effects of mixing on pulsational stability, we have tested schemes for determining the onset of convection that cover the likely range of convective core sizes for non-rotating stellar models---from Ledoux criterion to Schwarzschild criterion with convective overshoot. The qualitative conclusions of our stability analyses are insensitive to our choice of mixing parameters. Thus we present, as an illustrative example, models using the Schwarzschild criterion for convection and with convective overshoot of $10\%$ of one pressure scale height above the convective core \citep[with overshoot calculated using the prescription of][]{h00}. 



\begin{figure}
\centering
\includegraphics[width=\columnwidth]{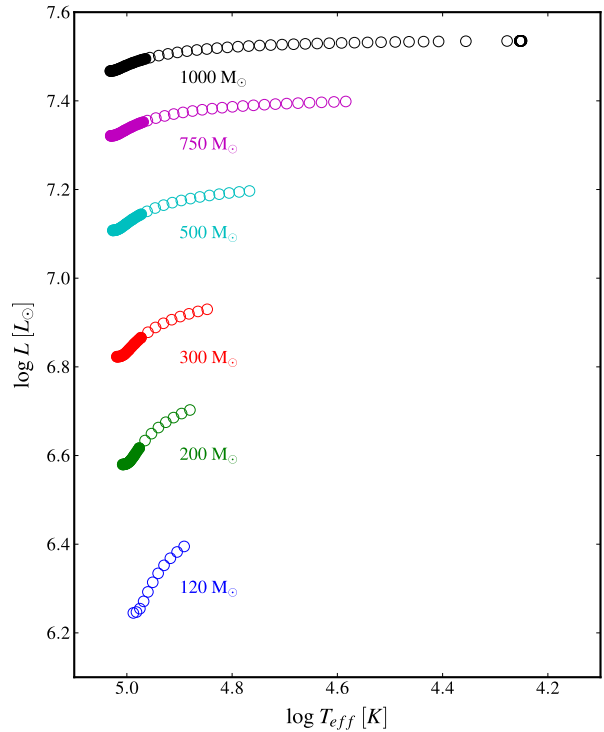}
\caption[H-R Diagram for massive Pop. III stars]{Hertzsprung-Russell Diagram for Population III Stars with masses between 120 and 1000 \Msun{} and initially primordial abundances. For each mass, the evolution is run with Schwarzschild mixing and 10\% convective overshoot, no rotation, and zero mass loss. Each sequence shown runs from the ZAMS to a center hydrogen composition ($X_{ctr}$) of 40\% by mass. Filled symbols indicate models with unstable radial pulsations when convective damping is ignored, while open symbols denote stable models. }\label{fig:hrd}
\end{figure}


We follow the evolution of Population III stars with masses between 120 and 1000 \Msun{} through their main sequence (core hydrogen burning) evolution, using the parameters outlined above and a nuclear reaction network covering all the relevant reactions for hydrogen and helium burning \citep[see][for details]{mesa11}. Figure \ref{fig:hrd} shows the evolution of these stellar models from the Zero-Age Main Sequence (ZAMS) to a 40\% central hydrogen mass fraction ($X_{ctr}$). For massive Population III stars ($M \gtrsim 20$ \Msun), hydrogen burning by the p-p chain cannot halt the gravitational contraction; the star contracts to $T_{c} \sim 10^{8} \; {\rm K}$ and ignites He fusion by the triple-alpha process to produce enough heavy elements to sustain hydrogen burning by the CNO cycle \citep{mgc+01}. Thus, these stars have much higher central temperatures on the main sequence than their higher metallicity counterparts.

The highly mixed models shown in Fig. \ref{fig:hrd} have large convective cores and correspondingly thin radiative envelopes, which expand to form red supergiants while the core is still fusing hydrogen. This is particularly true for the most massive models, as can be seen in Fig. \ref{fig:hrd}.\footnote{Evolving the most massive models to a lower central hydrogen mass fraction becomes difficult because the envelope expands significantly and evolves on a timescale much shorter than the nuclear timescale. For the less mixed models we have run (those without overshoot or using the Ledoux criterion for convection), the radiative envelopes are larger for a given mass, and they do not expand to form red supergiants while still burning hydrogen in their cores. Our qualitative results regarding the stability of main sequence massive stars are insensitive to our choice of mixing parameters. However, exactly when (and if) the star evolves significantly away from its ZAMS state does depend on the details of mixing.}

 



\section{Pulsational Analysis} \label{sec:pulse}

We assume the quasi-adiabatic approximation for our pulsational analysis. For each model in the sequences shown in Fig. \ref{fig:hrd} (and the analogous higher metallicity models), we have calculated a set of adiabatic oscillation modes using the recently released version of the \texttt{ADIPLS} oscillation package (\citeauthor[][{\it private communication}]{c08}). We assume zero Lagrangian pressure perturbation, $\delta p$, as our outer boundary condition for all mode calculations. In all our calculations below, we use $\delta x$ to represent the Lagrangian perturbation of variable $x$, and $x^{\prime}$ its Eulerian perturbation.

We calculate the perturbative driving and damping rates for the adiabatic modes following the formalism of \citet{uoa+89}. We thus approximate the imaginary component of the mode frequency (driving or damping rate) from the second-order expression for the entropy perturbation
\begin{equation} \label{eqn:deltaom}
	\delta \omega \equiv \Im\left(\omega\right) = \frac{1}{2} \frac{\intM 
		\frac{\delta T}{T}^{\ast} \delta \left(\epsnet - \frac{1}{\rho}
		\;\diverg{\vec{F}}\right) dm}{ \omega^{2} \intM \left|\delta r\right|^{2} dm}
\end{equation} 
The above represents the integration of the change in entropy due to the perturbation (mode) over one oscillation period, $2\pi / \omega$, and over the adiabatic mass, $M_{ad}$, of the star, defined as where the thermal time of the overlying mass is longer than the mode period \citep[cf.,][and our eq. \ref{eqn:nonad}]{n98}. In this formalism, positive (negative) imaginary values represent mode driving (damping). Here $\epsnet$ contains the energy generation due to nuclear processes, as well as cooling from neutrinos (negligible here) and gravitational power from contraction or expansion (negligible here).

We further separate the integral in the numerator into work integrals, as in \citet{uoa+89}, such that
\begin{align}
	\delta \omega =& \; \frac{1}{2} \frac{\Wnuc + \Wrad + \Wgrav - \dot{E}_{\rm conv}}
		{E_{mode}}\\
	E_{mode} =& \; \omega^{2} \, \intM \left|\delta r\right|^{2} dm\\ 
	\label{eqn:Wnuc}
	\Wnuc =& \intM \frac{\delta T}{T}^{\ast} \; \delta \epsnuc 
		\; dm\\ 
	\label{eqn:Wrad}
	\Wrad =& \intM \frac{\delta T}{T}^{\ast} \; \delta 
		\left(-\frac{1}{\rho} \; \diverg{\vec{F_{\rm rad}}}\right) 
		dm\\ 
	\label{eqn:Wgrav}
	\Wgrav =& \intM (\Gamma_1 - 1) (\Gamma_3 - 1) 
		\left|\frac{\delta \rho}{\rho}\right|^2 
		\epsilon_{g} \; dm\\ 
	&\epsilon_g = - \left(\epsilon_{\rm nuc} - 
  		\frac{1}{\rho} \nabla \cdot F\right)
\end{align}
Equation \ref{eqn:Wgrav} follows from \citet{ac75}, but never makes any significant contribution for the modes investigated here since $\epsilon_{g} \ll \epsnuc$ on the main sequence. Note that we also ignore the effect of neutrino losses to the damping of modes since, again, $\epsilon_{\rm neut} \ll \epsnuc$ on the main sequence. 

Below we provide more details on the meanings of each of these work integrals.


\subsection{Nuclear Driving} \label{subsec:eps}

Oscillation modes with large amplitudes near regions of nuclear burning in the stellar interior can couple to this energy generation for mode driving. Over one pulsation cycle, ignoring other effects, a mode will experience a net gain in energy due to the excess energy generated during the phase of positive temperature perturbation, due to the strong, positive temperature dependence of most nuclear processes, {\bf $\partial\ln{\epsilon} / \partial\ln{T}|_{\rho} \sim 10$} for hydrogen burning. 

Taking equation \ref{eqn:Wnuc}, and plugging in for $\delta \epsilon$ we find
\begin{equation}
	\Wnuc = \intM \left|\frac{\delta T}{T}\right|^{2} \; \left(\frac{\epsilon_{\rho, i}}{\Gamma_{3} - 1} + \epsilon_{T, i}\right) \epsilon_{i} \; dm,
\end{equation}
where the sum over consecutive indices, $i$, represents a sum over the nuclear reactions taking place. The factors $\epsilon_{T}$ and $\epsilon_{\rho}$ represent the logarithmic partial derivatives of the energy generation rate against $T$ and $\rho$ respectively. We have also used the fact that the Lagrangian perturbations are adiabatic to eliminate the density perturbation in favor of the temperature perturbation.

As pointed out by \citet{uoa+89} the values of $\epsilon_{T,\rho}$ are in general, frequency dependent. When the timescales associated with the individual reactions are sufficiently short or long compared to the mode period, simple approximations can be made to give appropriate, modified $\epsilon_{T,\rho}$ values \citep[as in the treatment of the p-p chain in][]{uoa+89}. However, when the timescales are comparable, one must account for phase shifts between the thermodynamic and abundance perturbations \citep[cf.,][]{k88,ss12}. 

For the high central temperatures reached in Population III stars ($\log T_{c} \gtrsim 8.1$), the fundamental mode period is intermediate between the proton capture and beta-decay timescales in the CNO cycle. Accounting for the phases of the abundance variations, we find a slight reduction of $\epsilon_{T}$ by $\sim 15\%$ from the equilibrium value determined by the temperature dependence of the $^{14}$N$(p,\;\gamma)^{15}$O reaction. Thus we have {\bf $\epsilon_{T} \lesssim 8$} near the centers of massive Population III stars, somewhat smaller than the value used in \citeauthor{bhw01}'s \citeyearpar{bhw01} earlier study.





\subsection{Radiative Damping} \label{subsec:raddamp}

Equation \ref{eqn:Wrad} can be rewritten as
\begin{equation} \label{eqn:WradExp}
	\Wrad = -\int_{0}^{R_{ad}} \frac{\delta T^{\star}}{T}\frac{d(\delta L_R)}{dr} dr,
\end{equation} 
where $L_{R}$ signifies the radiative luminosity, and $R_{ad}$ is the radius corresponding to the adiabatic mass, $M_{ad}$, introduced in equation \ref{eqn:deltaom} above. Note that, according to this expression, damping occurs whenever the temperature perturbation and radial derivative of the luminosity perturbation are of the same sign. That is to say, if, at the temperature maximum, there is more perturbed luminosity leaving the top of the layer than entering the bottom, there is radiative damping. 

The luminosity perturbation itself has several terms; keeping only those relevant for radial oscillations, it is given by
\begin{equation} \label{eqn:delL}
	\begin{split}
	\frac{\delta L_R}{L_R} =& \:\: 4\frac{\xi_r}{r} + 4\frac{\delta T}{T} - 
		\frac{\delta \kappa}{\kappa} - \frac{1}{\nabla_{\star}} 
		\left(1 - \frac{d\ln{\nabla_{ad}}}{d\ln{p}}\right) \frac{\delta T}{T} + \\ 
	& \frac{\nabla_{ad}}{\nabla_{\star}} \left[\left(\frac{4\pi r^3\rho}{m_r} - 
		\frac{\omega^2}{\omega_{\star}^2} \frac{x^3}{q} - 4\right)\frac{\xi_r}{r} 
		+ \frac{1}{g} \frac{d\Phi^\prime}{dr}\right],
	\end{split}
\end{equation} 
where $\omega_{\star}^{2}$ is the stellar natural frequency $G M_{\star} / R_{\star}^{3}$, $q$ is the local mass coordinate, $m_{r} / M_{\star}$; $x$ is $r / R_{\star}$; $\xi_{r}$ is the amplitude of radial displacement; $\nabla_{\star}$ is the logarithmic temperature derivative in the background model, $d\ln T / d\ln P$; $\nabla_{ad}$ is the adiabatic derivative, {\bf $(\partial \ln T / \partial\ln P)_{S}$}; $\kappa$ the opacity; $\rho$ the density; $g$ the local gravity; and $\Phi^{\prime}$ the Eulerian perturbation to the stellar potential.


For oscillation modes with significant amplitude in the stellar core, radiative diffusion generally leads to modest damping of modes. Radiative damping becomes increasingly important as mode energy shifts into the stellar envelope. At low enough effective temperatures and high enough metallicities, the opacity perturbation term ($\delta \kappa / \kappa$) can provide strong driving through the $\kappa$-mechanism by blocking flux at compression (maximum positive temperature perturbation) such that $d (\delta L_{R}) / dr$ changes sign \citep[e.g.,][]{c80}.

\subsection{Convective Damping} \label{subsec:convdamp}

The characteristic convective damping timescale for stars with convective cores is $\sim (M_{\star} R_{\star}^{2} / L_{\star} )^{1/3}$ \citep[e.g.,][]{z89}. For massive stars, this is $\sim 4$ orders of magnitude shorter than the thermal time, which is the characteristic growth time of perturbations due to the $\epsilon$-mechanism ($\kappa$-driving can give rise to much more rapid growth). This highlights the importance of including convective damping in calculations of the pulsational stability of massive stars. The subtlety is that the effective convective viscosity is significantly suppressed relative to the above estimate because of the mismatch between the convective turnover time and the mode periods of interest \citep{gn77,z89}.


\begin{figure*}
\centering
\includegraphics[width=\textwidth]{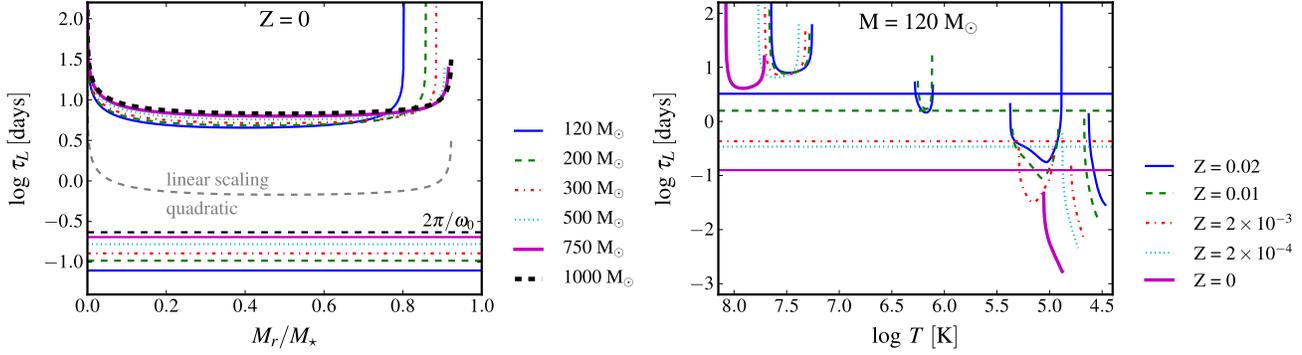}
\caption[Convective turnover time and fundamental mode periods]{Convective turnover time, $\tconv{}$, plotted against fractional mass enclosed for a Population III main sequence model with masses from $120 - 10^{3}$ \Msun{} (left panel), and against internal temperature for 120 \Msun{} models for a range of metallicities (right panel). For the Population III stars, the models shown have $X_{ctr} \approx 0.7$; the variable metallicity stars have $X_{ctr} \approx 0.4$. The straight horizontal lines in each panel show the fundamental mode periods corresponding to each model plotted. For reference, the boundary between linear and quadratic scaling of the convective viscosity with mode period for the 1000\Msun{} Population III model is shown as the thick dashed grey line in the left panel. For all other models (in both panels) this transition occurs at a period one order of magnitude less than the convective turnover time of the stellar model.}\label{fig:tauL}
\end{figure*}


Recent simulations by \citet{pbs09} have shown that, in agreement with the work of \citet{gn77} and \citet{z89}, interaction between pulsations and convective eddies indeed acts like a viscosity, damping oscillatory modes in convection zones. In particular, \citet{pbs09} find that, over the range of frequencies and spatial scales accessible in their simulations, the effective kinematic viscosity of the convection scales linearly with the ratio of the oscillation period and turnover time of the largest eddies. We refer to the latter as $\tconv$ below. 

However, their simulations do not allow resolution of eddies with turnover times smaller than about $0.3 \, \tconv$. That they see {\it no measurable viscous damping} for forcing periods less than this resolution limit implies that eddies with turnover times less than or equal to the forcing timescale dominate the viscous interaction when $t_{\rm force} \lesssim 0.3 \, \tconv$. Thus, these authors argue that the formalism of \citet{gn77}, derived on the assumption that near-resonant eddies in a Kolmogorov cascade dominate the viscosity, applies for higher frequency (shorter period) forcing. 

We thus take a conservative approach, in terms of minimizing the effect of viscous damping, and assume that the change to quadratic scaling occurs {\it just beyond the resolution of their simulations,} at a ratio of $t_{\rm force} / \tconv \equiv \Pimin = 0.1$. Generalizing the results of \citet{ps11}, we define our fiducial, frequency-dependent kinematic viscosity to be
\begin{equation} \label{eqn:convvisc}
	\nu\left(\omega, r\right) = \left(\frac{1}{3} v_{L} L\right) \min 
		\left[\frac{1}{\Pimin} \left|\frac{2\pi}{\omega \tconv}\right|^{2}, 
		\left|\frac{2\pi}{\omega \tconv}\right|, \Pimax\right].
\end{equation} 
Here $r$ is the radius of the layer, $L = \min(r,H)$ is the mixing length, $v_{L}$ is the convective velocity according to mixing length theory, $\tconv$ is the convective turnover time of the largest eddies (which we take to be $1 / N_{\rm Brunt-V\ddot{a}is\ddot{a}l\ddot{a}}$), $\omega$ is the mode's angular frequency (and $2\pi / \omega \equiv t_{\rm force}$), and the dimensionless factor $\Pimax = 2.4$ represents the timescale ratio above which the viscosity is independent of forcing period (saturation period). As can be seen, the quadratic reduction applies for $\left(2\pi / \omega \tconv\right) < \Pimin$, linear reduction for $\Pimin \leq \left(2\pi / \omega \tconv\right) \leq \Pimax$, and unreduced for $\left(2\pi / \omega \tconv\right) > \Pimax$ 


Figure \ref{fig:tauL} shows a comparison of $\tconv$ and the fundamental mode period for several main sequence models. The left panel shows Population III models, while the right shows higher metallicity, 120 \Msun{} models. For Population III models, the fundamental mode periods are always well into the quadratic scaling regime, but for the higher metallicity cases, the mode periods tend to lie in the linear scaling or unreduced regimes, especially for the convective zones in the envelope of the star ($\log T \lesssim 6$).

For a {\it radial mode} within the star, the integrated rate of convective damping is given by
\begin{equation}
	\dot{E}_{\rm conv} = \int \; \nu \, s_{0}^{\prime} \, \left(\partialderiv{v_{r}}{r}\right)^{2} dm.
\end{equation}
Plugging in the expansion in terms of modes, we find
\begin{equation} \label{eqn:Edotconv}
	\dot{E}_{\rm conv} = \int \; \nu \, s_{0}^{\prime} \, \omega^{2} \, \left|k_{r}\right|^{2} \, \xi_{r}^{2} \; dm,
\end{equation} 
where $s_{0}^{\prime}$ is the coefficient for damping due to the radial component of the shear from \citet[][their eq. 9]{ps11}, and $k_{r}$ is given by the usual adiabatic oscillation equations. In accordance with equation \ref{eqn:deltaom}, we then divide by the mode energy to calculate the integrated damping rate due to interaction with the convection. 

For the modes in Population III stars described below, $\left(2\pi / \omega \tconv\right) < \Pimin$. If the turnover to quadratic scaling actually occurs at mode periods shorter (longer) than what we have taken here, the damping will be stronger (weaker) than what we calculate. However, given the limiting value of $\Pimin \lesssim 0.3$ found by \citet{pbs09}, we find it unlikely that the damping is weaker than calculated here (see Sections \ref{sec:res} and \ref{sec:disc} for further discussion).  



\section{Results} \label{sec:res}

\subsection{Zero Metallicity} \label{ssec:z0res}


\begin{figure}
\centering
\includegraphics[width=3.5in]{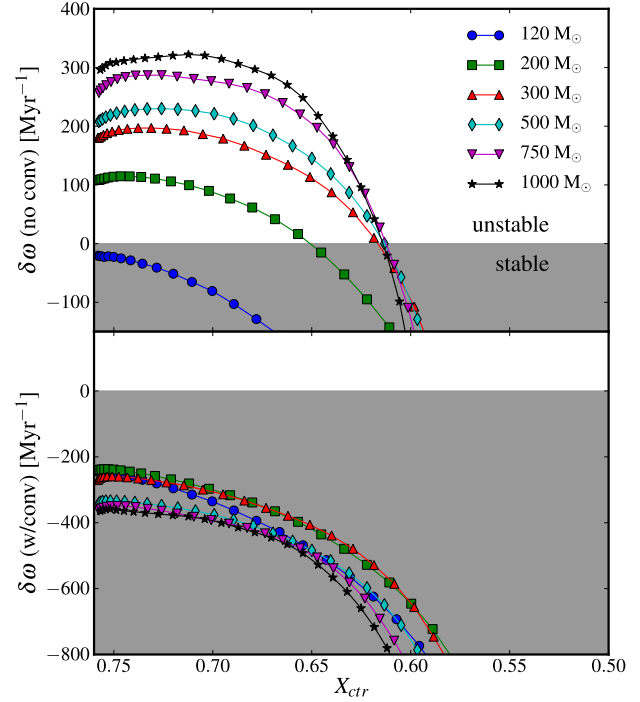}
\caption[Growth rates for modes in Pop III stars]{Growth rates, in units of Myr$^{-1}$, for the radial fundamental modes of Population III stars on the main sequence, shown versus the mass fraction of hydrogen in the star's convective core. The top panel shows the results assuming the ``frozen-in'' approximation for convection \citep[for comparison with fig. 4 of][]{bhw01}, while the bottom panel includes the effects of convective damping. Negative growth rates (shaded region) denote stability. Note that the 120 \Msun{} models are always stable in this analysis with or without convective damping, due to the lower CNO cycle temperature dependence (see Section \ref{subsec:eps})}\label{fig:rates}  
\end{figure}


Figure \ref{fig:rates} shows the results of our quasi-adiabatic stability analysis for the fundamental radial modes in Population III main sequence models. The top panel shows the driving/damping rates using the ``frozen-in'' approximation for convection, while the bottom panel includes the effects of convective damping. In the absence of convective damping, massive stellar models with $M > 120$ \Msun{} have unstable fundamental modes until the central hydrogen mass fraction is $\lesssim 0.6$. The bottom panel of Fig. \ref{fig:rates} demonstrates, however, that convective damping stabilizes the fundamental mode in all Population III main sequence models, regardless of mass and age. We find that this is true unless $\Pimin$ (which characterizes the transition from linear to quadratic suppression of convective damping) satisfies $\Pimin \gtrsim 0.25$.  While this possibility is not completely ruled out by Penev et al.'s calibration of convective damping, we regard it as exceptionally fine-tuned.  



When we ignore the effects of convective damping (top panel), our results are comparable to those of \citet[][their fig. 4]{bhw01} to within factor of 2. The moderate disagreement likely stems from differences in the parameters used to evolve the stellar models and the $\sim 15\%$ difference in the temperature dependence of the CNO cycle between their work and ours (see Section \ref{subsec:eps}). In particular, the reduction in temperature sensitivity leads us to conclude that the fundamental mode in 120 \Msun{} Population III main sequence models is stabilized by radiative diffusion in the envelope even before convective viscosity is taken into account.

While our calculations are quasi-adiabatic and theirs are non-adiabatic, this difference is likely unimportant for the modes and models shown in Fig. \ref{fig:rates}, which \citeauthor{bhw01} found to be unstable to the $\epsilon$-mechanism (stars having $X_{ctr} \gtrsim 0.5$). First, the non-adiabatic region of the star, defined as where 
\begin{equation} \label{eqn:nonad}
	\frac{\ttherm}{\tmode} \equiv \frac{\int_{m}^{M} c_{V}\; T\; dm}{\tmode} \lesssim 1
\end{equation} 
contains less than $\sim 10^{-4}$ of the total stellar mass for all models shown in Fig. \ref{fig:rates}. Second, the fraction of the mode energy found in the non-adiabatic zone is also always less than $\sim 10^{-4}$. Taken together, these justify our quasi-adiabatic assumption.



\begin{figure}
\centering
\includegraphics[width=3.5in]{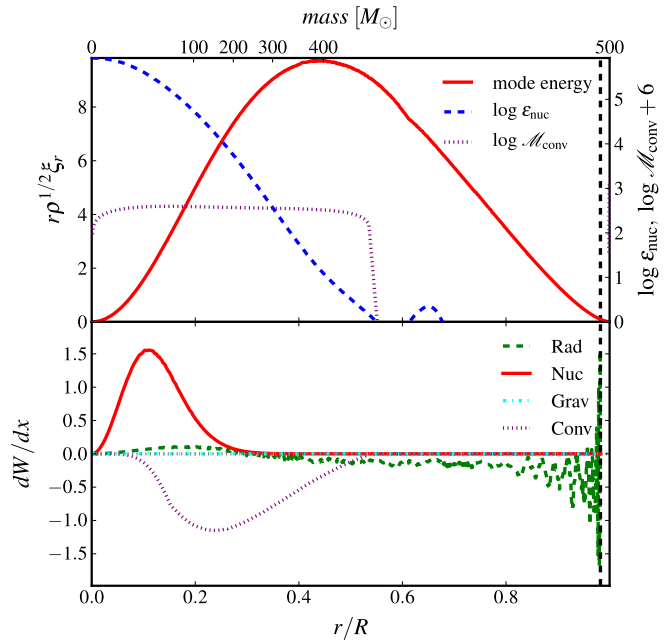}
\caption[Mode Work Integrals]{Mode energy (red solid line, top) and work integrals (bottom) for a typical, damped radial fundamental mode in a Population III, 500\Msun{} main sequence model. All quantities are plotted versus $r/R$ (bottom abcissa) and enclosed mass, $m_{r}$ (top abcissa). The top panel also shows the nuclear energy generation rate (blue dashed, right axis) and the Mach number of the convection (purple, dotted line). The bottom panel shows the different contributions to the total work integral described in equations \ref{eqn:Wnuc}, \ref{eqn:Wrad}, \ref{eqn:Wgrav}, and \ref{eqn:Edotconv}, with colors according to the legend.} \label{fig:work}
\end{figure}


Figure \ref{fig:work} shows the mode energy distribution (top panel) and work integrals (bottom panel) for the fundamental radial mode in a main sequence ($X_{ctr} = 0.718$) 500 \Msun{}, Population III stellar model. The dominant contributions to the total work integral come from the nuclear driving and convective damping, at $r/R \lesssim 0.4$ ($m_{r} \lesssim 100$ \Msun{}). Similar plots for fundamental modes in other main sequence models are qualitatively the same. For more evolved models (those with lower $X_{ctr}$), the mode energy shifts outward into the stellar envelope and radiative diffusion is the dominant source of damping.

%
%



\subsection{Dependence on Metallicity} \label{ssec:hiherZres}



\begin{figure}
\centering
\includegraphics[width=3.5in]{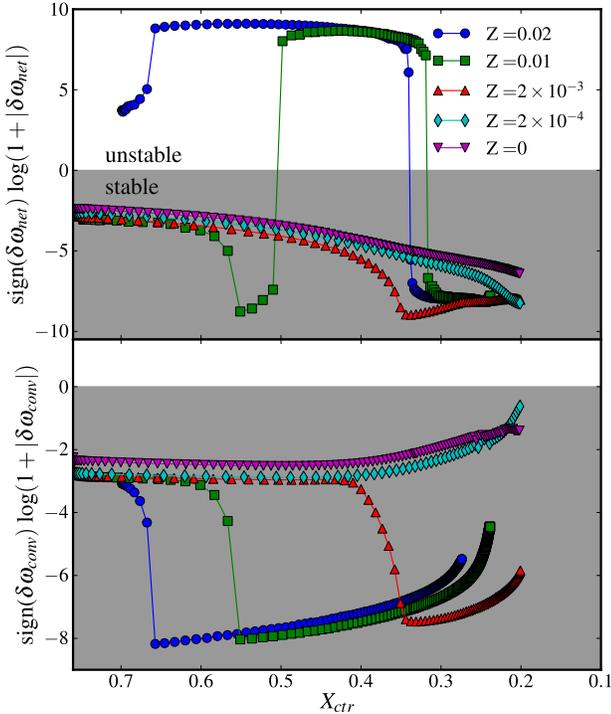}
\caption[Growth rates for modes in higher z stars]{Net growth rates including convective damping (top) and the convective damping rates (bottom) for radial fundamental modes in stars on the main sequence with $M = 120$ \Msun{} and a range of metallicities, shown versus the mass fraction of hydrogen in the star's convective core. Note the logarithmic scale of the y-axis, which highlights the large change in the magnitude of the driving/damping when the mode amplitude shifts to the envelope, as for the solar and half solar metallicity models (blue circles and green squares, respectively). The corresponding change in the magnitude of the convective damping occurs because the damping is dominated by an envelope convection zone rather than the convective core (see Fig. \ref{fig:tauL}).}\label{fig:Zrates}  
\end{figure}


We have performed the same analysis as for the Population III stars, for models with 120 \Msun{} and a range of metallicities from $10^{-6}$ to solar (taken to be $0.02$). Figure \ref{fig:Zrates} shows the results of these calculations. In the upper panel, the net growth rate (taking into account any convective damping) is shown versus the central hydrogen fraction (as in Fig. \ref{fig:rates}). Note the logarithmic scale for the growth and damping rates. The bottom panel shows the contribution of the convective viscosity to the damping rate. Models with $0 < Z < 2 \times 10^{-4}$ are not shown since their results are nearly identical to the $Z = 0$ models. 



\begin{figure*}
\centering
\includegraphics[width=\textwidth]{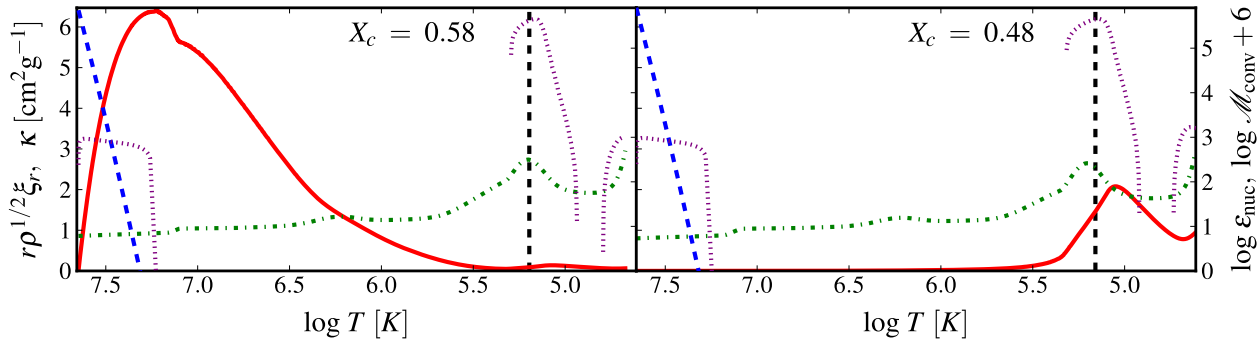}
\caption[Mode plots in Z = 0.01 models]{Mode energy (red) as a function of $\log T$ for $Z = 0.01$ main sequence stellar models. Shown on the left is a stable mode with $X_{ctr} \approx 0.58$, and on the right is a $\kappa$-driven mode with $X_{ctr} \approx 0.48$. Also shown are the run of opacity (green dot-dashed line, left ordinate, multiplied by 3 for clarity), convective mach number (purple dotted line, right ordinate), nuclear energy generation rate (blue dashed line, left ordinate), and adiabatic cutoff (black dashed vertical line). All values are shown versus $\log T$. In the more evolved model on the right, the low-density envelope has expanded by a factor of $\sim 2$ due to the enhanced radiation pressure around the iron opacity bump at $\log T \approx 5.2$.} \label{fig:z1m1modes}
\end{figure*}


As shown in Fig. \ref{fig:z1m1modes}, the sharp transition in the net growth and convective damping rates for models with $Z \gtrsim 2 \times 10^{-3}$ is due to the shift in mode energy out into the stellar envelope (shown in the figure for models with $Z = 0.01$). This transition occurs when the low-density envelope expands due to the increased radiation pressure around the iron opacity enhancement at $\log T \approx 5.2$. When the mode energy is primarily in the stellar envelope there is the possibility for $\kappa$-mechanism driving around the iron opacity enhancement, as well as stronger convective damping in an outer convection zone where the mode and convective frequencies are comparable. 
%


In agreement with \citet{bhw01}, we find that the radial fundamental modes in main sequence models with $Z \gtrsim 2 \times 10^{-3}$ are strongly driven by the $\kappa$-mechanism operating in the envelope at the location of the opacity-enhancement due (primarily) to iron transitions \citep{ri92}. The location of the adiabatic cutoff mass/radius, as defined by equation \ref{eqn:nonad}, relative to the location of the opacity peak at $\log T \approx 5.2$ determines when the opacity is capable of providing strong driving \citep[cf.][]{c80}. This alignment changes with the expansion of the stellar envelope, such that only models at particular evolutionary stages on the main sequence have fundamental modes driven by the $\kappa$-mechanism.




{\it Across the range of masses and metallicities investigated here, we find that convective viscosity always overcomes the $\epsilon$-driving in the stellar core.} However, for models with $Z \gtrsim 2 \times 10^{-3}$, the stellar and mode structures are such that $\kappa$-mechanism provides the dominant driving and is orders of magnitude stronger than the convective damping. However, this envelope driving occurs in a region where the local ratio $t_{\rm therm} / t_{\rm mode}$ is only slightly larger than unity and our assumption of quasi-adiabaticity is not entirely valid (as shown in Fig. \ref{fig:z1m1modes}). A fully-non-adiabatic method would likely provide more reliable growth-rates for these modes. Moreover, our quasi-adiabatic analysis is not capable of capturing strange-mode driving, which is also likely to be important for the same models that are unstable to the $\kappa$-mechanism \citep[cf.,][]{g98}.



%

\subsection{Higher order and non-radial modes} \label{ssec:nonrad}

In addition to the fundamental mode, we have also studied the stability of higher order radial modes and non-radial oscillations. Higher order radial modes---those having a larger number of radial nodes---have a more significant fraction of their energy outside the nuclear burning core, and have shorter wavelengths, which both lead to stabilization by radiative damping.





The presence of a large convective core prevents gravity(g)-modes from propagating to small radii and coupling to the nuclear driving \citep[see also][]{su11}. While high frequency, low angular degree pressure(p)-modes may penetrate into the burning region, these modes have their largest amplitudes near the stellar surface where the density is low and radiative damping dominates. Furthermore, while p- and g-modes of high enough angular degree can be trapped by the increase in $N^{2}$ at the edge of the convective core due to the chemical composition gradient, these modes experience strong radiative damping that overcomes the weak nuclear-driving they experience \citep[cf.][]{uoa+89}.



\section{Discussion and Conclusions} \label{sec:disc}

%
%

In this paper, we have studied the stability of radial and non-radial oscillation modes for massive main sequence stars of a range of metallicities, from primordial Population III stars to those of solar composition. We find that convective damping is likely to stabilize modes that would otherwise be unstable by the $\epsilon$-mechanism (driving by nuclear fusion). This is particularly important for primordial stars whose high effective temperatures imply that there is no driving due to opacity variations in the stellar envelope (the $\kappa$-mechanism). Thus, we conclude that massive Population III stars are linearly stable on the main sequence, supporting the idea that such stars are unlikely to experience any pulsation-driven mass loss during their main sequence evolution.


The calculations in this paper build on those of \citet{bhw01} and \citet{su11}, who found that the fundamental radial oscillations in Population III stars with $M > 120$ \Msun{} are unstable due to the $\epsilon$-mechanism. However, both \citeauthor{bhw01} and \citeauthor{su11} neglected convective damping. Recent hydrodynamic simulations by \citet{pbs09} have shown that the convective viscosity scales linearly with the oscillation period for periods of the same order as the turnover time of the largest eddies \citep[in agreement with][]{z89} and quadratically for much shorter periods \citep[in agreement with][]{gn77}. The transition between these regimes is not directly accessible to the simulations; we have taken a conservative estimate of $\Pimin = 0.1$, in agreement with the simulations, in order to minimize the convective viscosity. If $\Pimin \gtrsim 0.25$, some $\epsilon$-driven modes remain unstable, but this is an extremely fine-tuned value, given \citeauthor{pbs09}'s result that $\Pimin \lesssim 0.3$. 

Even more recent simulation work by \citet{ol12} calls the work of \citeauthor{pbs09} into question, finding that the viscosity always scales with the period ratio squared \citep[as][]{gn77}, and may in fact be negative (i.e., driving) in the short period forcing regime of interest for this work. These latter simulations use a smaller forcing amplitude than that of \citeauthor{pbs09}, leading to relatively noisier measurements, and a smaller simulation box which may not capture the full spectrum of turbulence on the large scales within a stellar convection zone. This latter aspect may be crucial to their measurement since they infer that the negative contribution to the viscosity is dominated by the effect of the largest scale eddies. The sensitivity of our results to the exact form of the convective viscosity highlights the need for higher resolution simulations of the convection-oscillation interaction, particularly using larger simulation domains to capture the interaction with the large-scale eddies.

Stellar models with sufficiently high metallicity, $Z \gtrsim 2 \times 10^{-3}$, have strong opacity variations in their envelopes which can provide $\kappa$-mechanism driving. We find that the fundamental mode in these models are destabilized by the $\kappa$-mechanism operating near the iron-bump in opacity for much of their main sequence evolution. This driving is orders of magnitude stronger than the convective damping. These models experience large linear growth rates that approach non-adiabiticity ($\delta \omega \lesssim \omega$). For lower metallicity models, the $\epsilon$-mechanism provides the dominant driving but is overcome by convective damping as in Population III models. In all of the massive stellar models we have considered, we find that damping due to convective viscosity is stronger than the driving by the $\epsilon$-mechanism in stellar interiors for radial modes. 



Both \citeauthor{bhw01} and \citeauthor{su11} estimate that massive Population III stars may lose a few (for $M \lesssim 1000$ \Msun{}) or up to $\sim 10\%$ (for $M \gtrsim 1000$ \Msun{}) of their total stellar mass due to pulsation-driven mass loss during the proposed unstable phase of the main sequence. As \citeauthor{su11} highlight, the convective cores of stars $\gtrsim 500$ \Msun{}  comprise $\gtrsim 90\%$ of the stellar mass, implying that pulsation-driven mass loss may expose the convective core, perhaps further increasing mass loss by mixing nuclear-processed material to the surface to initiate a radiation-driven wind, or by increasing the duration of the proposed pulsationally unstable phase. If, instead, the pulsations are damped by convective viscosity as we have concluded, there may be no significant mass lost during the main sequence phase for massive Population III stars.

In order to fully address the mass loss properties of massive primordial stars, we must address their stability at more evolved stages. However, as stars evolve off the main sequence, their fundamental radial mode becomes increasingly non-adiabatic as the mode energy shifts to the expanding, low-density stellar envelope. Analyzing these modes requires a non-adiabatic calculation, which we leave to future work.
 

With all oscillatory modes in main sequence Population III stars likely stabilized by convective damping, evidence points to there being no mass loss on the main sequence for these metal-free stars, so long as the non-rotating stellar models used here are a reasonable approximation. The question of what happens as these stars evolve thus becomes all the more crucial for assessing the fates of the first stars.

\section*{Acknowledgments}

We thank Gordon Ogilvie for useful discussions regarding calibration of the convective viscosity. We also thank the anonymous referees for useful comments. This work was supported by NASA Headquarters under the NASA Earth and Space Science Fellowship Program - Grant 10-Astro10F-0030. 

\bibliography{starsrefs}

\label{lastpage}


\end{document}